\documentclass[journal]{IEEEtai}

\usepackage[colorlinks,urlcolor=blue,linkcolor=blue,citecolor=blue]{hyperref}

\usepackage{color,array}
\usepackage{algorithmic}
\usepackage{graphicx}
\usepackage{amsmath}
\usepackage{algorithm}
\usepackage{multirow}
\usepackage{subfig}

\begin{document}

\title{PrivMVMF: Privacy-Preserving Multi-View Matrix Factorization for Recommender Systems}

\author {Peihua Mai, \IEEEmembership{Member, IEEE}, Yan Pang, \IEEEmembership{Member, IEEE}, 
\thanks{This paragraph of the first footnote will contain the date on which you submitted your paper for review.}
\thanks{Yan Pang and Peihua Mai are with the Department of Analytics and Operations, National University of Singapore, 119077 Singapore (e-mail: jamespang@nus.edu.sg; peihua.m@u.nus.edu).}
}


\maketitle

\begin{abstract}
With an increasing focus on data privacy, there have been pilot studies on recommender systems in a federated learning (FL) framework, where multiple parties collaboratively train a model without sharing their data. Most of these studies assume that the conventional FL framework can fully protect user privacy. However, there are serious privacy risks in matrix factorization in federated recommender systems based on our study. This paper first provides a rigorous theoretical analysis of the server reconstruction attack in four scenarios in federated recommender systems, followed by comprehensive experiments. The empirical results demonstrate that the FL server could infer users' information with accuracy $>80\%$ based on the uploaded gradients from FL nodes. The robustness analysis suggests that our reconstruction attack analysis outperforms the random guess by $>30\%$ under Laplace noises with $b\leq 0.5$ for all scenarios. Then, the paper proposes a new privacy-preserving framework based on homomorphic encryption, Privacy-Preserving Multi-View Matrix Factorization (PrivMVMF), to enhance user data privacy protection in federated recommender systems. The proposed PrivMVMF is successfully implemented and tested thoroughly with the MovieLens dataset. 
\end{abstract}

\begin{IEEEImpStatement}
The recommender system is one of the most successful and popular AI applications in the industry. Multi-View Matrix Factorization (MVMF) has been proposed as an effective framework for the cold-start recommendation, where side information is incorporated into matrix factorization. MVMF requires users to upload their personal data to a centralized recommender, which raises serious privacy concerns. A solution to address the privacy problem is federated learning. However, the conventional federated MVMF allows the exchange of plaintext gradients, which is susceptible to information leakage. There is a lack of thorough study of the privacy risk in conventional federated MVMF in previous research works. To fill this gap, this paper provides rigorous theoretical analysis and comprehensive experiments on the privacy threat in the traditional federated MVMF. The analysis and experiments demonstrate that the server could reconstruct users' original information even when small amounts of noise are added to the gradients. Then we propose a PrivMVMF framework to address the information leakage problem successfully.  
\end{IEEEImpStatement}


\section{Introduction}
\label{intro}
\IEEEPARstart{T}{he} recommendation system relies on collecting users' personal information, such as purchase history, explicit feedback, social relationship, and so on. Recently, some laws and regulations have been enacted to protect user privacy, which places constraints on the collection and exchange of users' personal data.

To protect user privacy, one way is to develop a recommendation system in federated learning (FL) framework that enables the clients to jointly train a model without sharing their data. In the FL setting, each client computes the updated gradient locally and sends the model update instead of the original data to a central server. The server then aggregates the gradients and updates the global model \cite{kairouz2019advances}. 

Collaborative filtering (CF) is one of the most effective approaches in recommendation systems \cite{su2009survey}, and matrix factorization (MF) is a popular technique in CF algorithms. MF decomposes a user-item interaction matrix into two low-rank matrices: user latent factors and item latent factors, which are used to generate the preference prediction \cite{koren2009matrix}. One disadvantage of MF-based recommendation is the cold-start problem: if an item or user has no rating information, the model cannot generate a latent factor representation for it and thus suffers difficulty in performing MF recommendation. A solution to the cold-start issue is to incorporate side information, i.e., user and item attributes, into matrix factorization.

Various approaches have been proposed for centralized recommender systems \cite{cortes2018cold, singh2008relational, seroussi2011personalised, gantner2010learning}. However, few studies have researched the topic in the federated setting. To the best of our knowledge, Flanagan et al. \cite{flanagan2020federated} is the first to propose a federated multi-view matrix factorization (MVMF) to address this problem. However, this method assumed that the conventional FL framework could fully protect user privacy. However, severe privacy risks exist in the federated MVMF recommender system, which is susceptible to server reconstruction attacks, i.e., the attack to recover users' sensitive information.  

To fill this gap, this paper first provides a theoretical analysis of the privacy threat of the federated MVMF method. In theoretical analysis, we develop server reconstruction attacks in four scenarios based on different treatments on unobserved ratings and methods to update user latent factors. The empirical study results indicate that the original federated MVMF method could leak users' personal information. Then, we design a privacy-preserving federated MVMF framework using homomorphic encryption (HE) to enhance the user data privacy protection in federated recommender systems. 

The main contributions of this paper are twofold:

(1) To the best of our knowledge, we are the first to provide a rigorous theoretical analysis of server reconstruction attacks in the federated MVMF recommender system. We also conducted comprehensive experiments, which show that the server could infer users' sensitive information with accuracy $>80\%$ using such attacks, and the attack is effective under a small amount of noise.

(2) To overcome the information leakage problem, we propose \textit{PrivMVMF}, a privacy-preserving federated MVMF framework enhanced with HE. The proposed framework has two advantages: a) To balance the tradeoff between efficiency and privacy protection, it adopts a strategy in which some unrated items are randomly sampled and assigned a weight on their gradients. b) To reduce complexity, it allocates some decrypting clients to decrypt and transmit the aggregated gradients to the server. A prototype of \textit{PrivMVMF} is implemented and tested on the movielens dataset.

\section{Literature Review}
\label{literature}
\textbf{Federated Matrix Factorization:} Federated recommender systems enable parties to collaboratively train the model without putting all data on a centralized server. Several federation methods for recommender systems have been introduced in recent works. Ammad-ud-din et al. \cite{ammad2019federated} proposed a federated matrix factorization method for implicit feedback. Each client updates the user latent factor locally and sends back item latent factor gradient to the server for aggregation and update. Duriakova et al. \cite{duriakova2019pdmfrec} presented a decentralized approach to matrix factorization without a central server, where each user exchanges the gradients with their neighbors. Lin et al. \cite{lin2020meta} provided a matrix factorization framework based on federated meta learning by generating private item embedding and rating prediction model. The above works haven't considered cold-start recommendation. To address the problem, Flanagan et al. \cite{flanagan2020federated} devised a federated multi-view matrix factorization based on the implicit feedback (e.g., clicks), where three matrices are factorized simultaneously with sharing latent factors.

\textbf{Cryptographic Techniques in Federated Recommender System:} Some studies used encryption schemes to develop privacy-preserving recommendation systems. Chai et al. \cite{chai2020secure} introduced FedMF, a secure federated matrix factorization framework. To increase security, each client can encrypt the gradient uploaded to the server with HE. Shmueli et al. \cite{shmueli2017secure} proposed multi-party protocols for item-based collaborative filtering of vertical distribution settings. In the online phase, the parties communicate only with a mediator that performs computation on encrypted data, which reduces communication costs and allows each party to make recommendations independent of other parties. Although both \cite{chai2020secure} and our paper adopt HE to enhance the security, our work extends the method by introducing decrypting clients and sampling of unrated items. The decrypting clients improve the efficiency to perform parameters updates, and the unrated items sampling strikes a balance between efficiency and privacy protection.

To the best of our knowledge, Flanagan et al. \cite{flanagan2020federated} is the first to devise a federated multi-view matrix factorization to address the cold-start problem, where the users directly upload the plaintext gradients to the server, and no work has considered the information leakage from the gradients. This paper first demonstrates the feasibility of server reconstruction attack, and then proposes a framework to enhance privacy protection. The study is conducted based on the assumption of honest clients and an honest-but-curious server \cite{yang2019federated}.

\section{Federated MVMF}
\label{fedmvmf}
The federated MVMF proposed by Flanagan et al. \cite{flanagan2020federated} is based on implicit feedback. In this section, we extend the framework to explicit feedback.
\subsection{Notations}
Table \ref{notation} lists the notations and their descriptions used throughout this paper.

\begin{table*}[ht]
\caption{Notations Used in the Paper}
\vskip 0.1in
\centering
\begin{tabular}{cccc}
\hline
\textbf{Notation} & \textbf{Description} & \textbf{Notation} & \textbf{Description}\\
\hline
$n$ & The number of users & $V,\ v_{d_y}$ & Item feature latent factor \\
$m$ & The number of items & $c$ & Uncertainty coefficient \\
$l_x$ & Dimension of user attributes & $K$ & Dimension of latent factor \\
$l_y$ & Dimension of item attributes & $\lambda$ & Regularization coefficient \\
$R,\ r_{i,j}, \hat{r}_{i,j}$ & Rating matrix &  $O_i$ & Set of rated items for user $i$\\
$X,\ x_{i,d_u}$ & User feature & $Y,\ y_{j,d_y}$ & Item feature \\
$U,\ u_{d_u}$ & User feature latent factor & $\gamma$ & Learning rate \\
$P,\ p_i$ & User latent factor & $\beta_1,\ \beta_2$ & Exponential decay rate \\
$Q,\ q_j$ & Item latent factor & $\varepsilon$ & Small number \\
\hline
\end{tabular}
\label{notation}
\vskip 0.1in
\end{table*}

\subsection{Multi-view Matrix Factorization}
Multi-view matrix factorization is performed on the three data sources: the rating matrix $R_{n\times m}$, the user attribute matrix $X_{n\times l_x}$, and the item content matrix $ Y_{m\times l_y}$, for $n$ users with $l_x$ features, and m items with $l_y$ features. The decomposition of the three matrices is given as:
\begin{equation}
R\approx PQ^T,X \approx PU^T,Y\approx QV^T
\end{equation}
where ${P=P}_{n\times K}$, ${Q=Q}_{m\times K}$, ${U=U}_{l_x\times K}$, ${V=V}_{l_y\times K}$ with $K$ representing the number of latent factors. For $P$ and $Q$, each row represents the latent factors for each user and item respectively. For $U$ and $V$, each row represents the latent factors for each feature of user and item respectively. The predicted rating of user u on item i is given as:
\begin{equation}
{\hat{r}}_{u,i}=p_u^Tq_i
\end{equation}

The latent factor representation is learned by minimizing the following cost function:
\begin{equation}
\begin{gathered}
J=\sum_{i}\sum_{j}{c_{i,j}(r_{i,j}-p_iq_j^T)^2}\\
+\lambda_1(\sum_{i}\sum_{d_u}(x_{i,d_u}-p_iu_{d_u}^T)^2+\sum_{j}\sum_{d_y}(y_{j,d_y}-q_jv_{d_y}^T)^2)\\
+\lambda_2(\sum_{i}||p_i||^2+\sum_{j}||q_j||^2+\sum_{d_u}||u_{d_u}||^2+\sum_{d_y}||v_{d_y}||^2)
\end{gathered}
\end{equation}
where $\lambda_1$ is used to adjust how much information the model should learn from side data, and $\lambda_2$ is a regularization term to prevent overfitting. $r_{i,j}=0$ if the rating is unobserved, and $r_{i,j}>0$ otherwise. $c_{i,j}$ could be treated as a weight on the error term for each rating record. This paper considers two definitions of $c_{i,j}$:
\begin{itemize}
    \item \textit{ObsOnly:} $c_{i,j}=1$ if $r_{i,j}>0$, and $c_{i,j}=0$ if $r_{i,j}=0$. Then the loss function only minimize the square error on the observed ratings.
    \item \textit{InclUnc:} $c_{i,j}=1$ if $r_{i,j}>0$, and $c_{i,j}=\alpha$ if $r_{i,j}=0$, where $0<\alpha<1$ is an uncertainty coefficient on the unobserved ratings. This case assigns a lower weight on the loss for unobserved ratings.
\end{itemize}
The matrix factorization for explicit feedback typically employs the first definition to reduce the bias of unobserved interaction and improve efficiency. However, employing the second definition reveals less information to the FL server. Furthermore, as is shown in section \ref{serveattack} and \ref{experiment}, adopting the second definition would present a challenge for the server attack. Therefore, we will consider both cases when designing the server attack.

\subsection{Federated Implementation}
The federated setting consists of three parties: clients, FL server, and item server. Each client holds their ratings and attributes locally and performs local update of $P$. FL server receives the gradients from clients and item server, and updates $U$ and $Q$. Item server is introduced to facilitate the training process. It stores the item features and conducts update of $V$. The following explains the details in the updates of each latent factor matrix.

\textbf{User feature latent factor $U$} is updated on the FL server with the formula as:
\begin{equation}
\label{updateU}
u_{d_u}^t=u_{d_u}^{t-1}-\gamma\frac{\partial J}{\partial u_{d_u}}
\end{equation}
where:
\begin{equation}
\begin{gathered}
\label{Ugrad}
\frac{\partial J}{\partial u_{d_u}}=-2\sum_{i}f(i,d_u)+2\lambda_2u_{d_u}
\end{gathered}
\end{equation}
where $f(i,d_u)=\left(x_{i,d_u}-p_iu_{d_u}^T\right)p_i$ is computed on each user locally.

\textbf{Item latent factor $Q$} is updated on the FL server with the formula as:
\begin{equation}
\label{updateQ}
q_j^t=q_j^{t-1}-\gamma\frac{\partial J}{\partial q_j}
\end{equation}
where:
\begin{equation}
\label{Qgrad}
\begin{gathered}
\frac{\partial J}{\partial q_j}=-2\sum_{i}f(i,j)
-2\lambda_1\sum_{d_y}f(j,d_y)+2\lambda_2q_j
\end{gathered}
\end{equation}
where $f(j,d_y)=(y_{j,d_y}-v_{d_y}q_j^T)v_{d_y}$ is computed on the item server, and $f(i,j)=c_{ij}(r_{i,j}-p_iq_j^T)p_i$ is computed on each user locally. Noted that for \textit{ObsOnly}, if $c_{ij}=0$, the user only computes and sends the gradients of items with $c_{ij}>0$, i.e., the rated items. For \textit{InclUnc}, the gradients for all items will be sent to the server.

Both user latent factor $P$ and item feature latent factor $V$ adopt two updating methods: 
\begin{itemize}
\item Semi-Alternating Least Squares (\textit{SemiALS}): Optimal $P$ and $V$ are computed using closed form formula under fixed $U$ and $Q$. Other parameters are updated using gradient descent method.
\item Stochastic Gradient Descent (\textit{SGD}): All of the parameters are updated using gradient descent method.
\end{itemize}
The time complexity for \textit{SemiALS} is $O(mK^2+K^3)$ per iteration, higher than that of \textit{SGD}. However, \textit{SGD} requires more iterations to achieve the optimum. \cite{yu2014parallel}

\textbf{User latent factor $P$} is updated on each client locally. For \textit{SemiALS}, it's updated with the formula as:
\begin{equation}
\label{pals}
p_i^\ast=(r_iC^{(i)}Q+\lambda_1x_iU)(Q^TC^{(i)}Q+\lambda_1U^TU+\lambda_2I)^{-1}
\end{equation}
where $C^{(i)}$ is a $m\times m$ diagnal matrix with $C_{jj}^{(i)}=c_{i,j}$.

For \textit{SGD}, it's updated with the formula as:
\begin{equation}
\label{updatepSGD}
p_i^t=p_i^{t-1}-\gamma\frac{\partial J}{\partial p_i}
\end{equation}
where:
\begin{equation}
\begin{gathered}
\label{pgrad}
\frac{\partial J}{\partial p_i}=-2\sum_{i}{c_{ij}(r_{i,j}-p_iq_j^T)q_j}\\
-2\lambda_1\sum_{d_u}{(x_{i,d_u}-p_iu_{d_u}^T)u_{d_u}}+2\lambda_2p_i
\end{gathered}
\end{equation}

\textbf{Item feature latent factor $V$} is updated on the item server. For \textit{SemiALS}, it's updated with the formula as:
\begin{equation}
\label{updateV}
    v_{d_y}^\ast=(y_{d_y}Q)(Q^TQ+\frac{\lambda_2}{\lambda_1}I)^{-1}
\end{equation}

For \textit{SGD}, it's updated with the formula as:
\begin{equation}
v_{d_y}^t=v_{d_y}^{t-1}-\gamma\frac{\partial J}{\partial v_{d_y}}
\end{equation}
where:
\begin{equation}
\begin{gathered}
\frac{\partial J}{\partial v_{d_y}}=-2\sum_{i}{(y_{i,d_v}-q_jv_{d_v}^T)q_j}+2\lambda_2v_{d_v}
\end{gathered}
\end{equation}

Algorithm \ref{alg:FedMVMF} outlines the federated implementation of MVMF (\textit{FedMVMF}). The gradient descent of U and Q are performed using Adaptive Moment Estimation (Adam) method to stabilize the convergence. 

\begin{algorithm}[htp]
   \caption{\textit{FedMVMF}}
   \label{alg:FedMVMF}
\begin{algorithmic}
\STATE \textbf{FL Server:}
\STATE \textbf{Initialize} $U$ and $Q$.
\FOR{t = 1 \textbf{to} T}{
\STATE Receive and aggregate $f(i,j)$ and $f(i,d_u)$ 
from user $i$ for $i\in[1,\ n]$. 
\STATE Receive $f(j,d_v)$ from item server. 
\STATE Update $U$ using equation (\ref{updateU}).
\STATE Update $Q$ using equation (\ref{updateQ}).
}
\ENDFOR
\STATE
\STATE \textbf{Item Server:}
\WHILE{True}{
\STATE Receive $Q$ from FL server.
\STATE Compute local $V$ using equation (\ref{updateV}).
\STATE Compute item latent factor gradients
$f(j,d_v)$.
\STATE Transmit gradients to server.
}
\ENDWHILE
\STATE
\STATE \textbf{Client:}
\WHILE{True}{
\STATE Receive $U$ and $Q$ from server.
\STATE Compute local $p_i$ using equation (\ref{pals}).
\STATE Compute $U$ gradients
$f(i,d_u)$ for $d_u\in[1,l_x]$.
\STATE Compute $Q$ gradients
$f(i,j)$ for $j\in[1,m]$.
\STATE Transmit gradients to server.
}
\ENDWHILE
\end{algorithmic}
\end{algorithm}

\subsection{Cold-start recommendation}
The recommendation for new users and items is discussed as followed.

\textbf{Cold-start user recommendation:} for any new user $i$, the system first generates the user latent factor $p_i$ based on the user’s attribute $x_i$ and the user feature latent factor matrix $U$. Then the predicted rating of user $i$ on item $j$ is given by the inner product of $p_i$ and $q_i$. $p_i$ is calculated by minimizing the loss function:
\begin{equation}
    J=\lambda_1\sum_{d_u}(x_{i,d_u}-p_iu_{d_u}^T)^2+\lambda_2\sum_{d_u}||u_{d_u}||^2
\end{equation}

The optimal solution of $p_i$ is defined as:
\begin{equation}
p_i^\ast=x_iU(U^TU+\frac{\lambda_2}{\lambda_1}I)^{-1}
\end{equation}

\textbf{Cold-start item recommendation:} given a new item $j$, the system first generates the user latent factor $q_j$ based on the item’s feature $y_j$ and the item feature latent factor matrix $V$. The estimated $q_i$ is then used to compute the predicted rating. $q_j$ is calculated by minimizing the loss function:
\begin{equation}
    J=\lambda_1\sum_{d_y}(y_{j,d_y}-q_jv_{d_y}^T)^2+\lambda_2\sum_{d_y}||v_{d_y}||^2
\end{equation}

The optimal solution of $q_j$ is defined as:
\begin{equation}
   q_j^\ast=y_jV(V^TV+\frac{\lambda_2}{\lambda_1}I)^{-1}
\end{equation}

\section{Sever Reconstruction Attack Analysis}
\label{serveattack}
In \textit{FedMVMF}, the FL server could reconstruct the user ratings and attributes based on the gradients they received. In this section, we consider the attacks for both \textit{SemiALS} and \textit{SGD} updates on user latent factor. Within each case, the attacks are slightly different between \textit{ObsOnly} and \textit{InclUnc}. The analysis is based on the assumption of honest clients and an honest-but-curious server.

\subsection{Reconstruction Attack for \textit{SemiALS} Update}
For \textit{SemiALS} the FL server is able to recover the user information within only one epoch given that the server has access to $U$ and $Q$.

\textbf{Attack for \textit{ObsOnly}:} In this case, the clients only upload the gradients for items with observed ratings. Therefore, for any user $i$, the gradients which the FL server receives is given by:
\begin{equation} \label{gradALSObsOnly}
\begin{gathered}
f(i,j)=(r_{i,j}-p_iq_j^T)p_i, j \in O_i\\
f(i,d_u)=(x_{i,d_u}-p_iu_{d_u}^T)p_i, d_u \in [1, l_x]\\
\end{gathered}
\end{equation}
where $f(i,j)$ and $f(i,d_u)$ denote the vector of gradient with length $K$, $O_i$ denotes the collection of items rated by user $i$, and $l_x$ denote the number of user attributes.

In \textit{SemiALS}, $p_i$ is updated by equation (\ref{pals}). Given that $c_{i,j}=0$ when $r_{i,j}=0$, the formula could be reduced to:
\begin{equation}
    p_i=(r_iQ_i+\lambda_1x_iU)(Q_i^TQ_i+\lambda_1U^TU+\lambda_2I)^{-1}
\end{equation}
where $r_i$ is the vector of observed ratings, and $Q_i=Q_{|O_i|\times K}$ is the latent factors for items rated by user $i$.

Let $A_i = Q_i(Q_i^TQ_i+\lambda_1U^TU+\lambda_2I)^{-1}$, and $B_i=\lambda_1U(Q_i^TQ_i+\lambda_1U^TU+\lambda_2I)^{-1}$, both of which could be computed on the FL server. Then $p_i$ could be written as $p_i=r_iA_i + x_i B_i$. Plugging into equation (\ref{gradALSObsOnly}), we have:
\begin{equation}
 \left\{
\begin{array}{l}
\begin{gathered}
f^Q_i=r_i^Tr_iA_i+ r_i^Tx_iB_i- Q_i(A_i^TR_rA_i\\+A_i^TR_xB_i+B_i^TX_rA_i+B_i^TX_xB_i)\\
f^U_i=x^T_ir_iA_i+ x^T_ix_iB_i- U(A_i^TR_rA_i\\+A_i^TR_xB_i+B_i^TX_rA_i+B_i^TX_xB_i)\\
\end{gathered}
\end{array}
\right.
\end{equation}
where $f^Q_i={f^Q_i}_{|O_i|\times K}$ with $j^{th}$ row being $f(i,j)$, $f^U_i={f^U_i}_{l_x\times K}$ with $d_u^{th}$ row being $f(i,d_u)$, and:
\begin{equation}
\begin{gathered}
    R_r=r_i^Tr_i, R_x=r_i^Tx_i, \\
    X_r=x^T_ir_i, X_x=x^T_ix_i
\end{gathered}
\end{equation}

Then the FL server obtain a second order non-linear system with $(l_x+|O_i|)\times K$ equations, consisting of $(l_x+|O_i|)$ variables, $r_i$ and $x_i$. Therefore, it's plausible to find the solution of user ratings $r_i$ and user attributes $x_i$ using methods such as Newton-Raphson algorithm. To reconcile the number of equations and variables, we choose a random factor $n \in [1,K]$, and solve the equation systems under the fixed $n$.

\textbf{Attack for \textit{InclUnc}:} In this case the client sends gradients of all items to the FL server, multiplied by a uncertainty coefficient $c_{i,j}$. For any user $i$, the gradients the FL server is given by:
\begin{equation} \label{gradALSInclUnc}
\begin{gathered}
f(i,j)=c_{i,j}(r_{i,j}-p_iq_j^T)p_i, j \in [1, m]\\
f(i,d_u)=(x_{i,d_u}-p_iu_{d_u}^T)p_i, d_u \in [1, l_x]\\
\end{gathered}
\end{equation}

 Let $A_i' = C^{(i)}Q(Q^TC^{(i)}Q+\lambda_1U^TU+\lambda_2I)^{-1}$, and $B_i'=\lambda_1U(Q^TC^{(i)}Q+\lambda_1U^TU+\lambda_2I)^{-1}$. Then $p_i$ can be written as:
\begin{equation}
\begin{gathered}
    p_i = r_iA_i'+x_iB_i'
\end{gathered}
\end{equation}
Plugging into equation (\ref{gradALSInclUnc}), we can obtain the final equation system:
\begin{equation}\label{finalALSInclUnc}
 \left\{
\begin{array}{l}
\begin{gathered}
f^Q_i=C^{(i)}\bigl(r_i^Tr_iA_i'+ r_i^Tx_iB_i- Q((A_i')^TR_rA_i'\\+(A_i')^TR_xB_i'+(B_i')^TX_rA_i'+(B_i')^TX_xB_i')\bigr)\\
f^U_i=x^T_ir_iA_i'+ x^T_ix_iB_i'- U((A_i')^TR_rA_i'\\+(A_i')^TR_xB_i'+(B_i')^TX_rD_i+(B_i')^TX_xB_i')\\
\end{gathered}
\end{array}
\right.
\end{equation}
where $r_i$ is user $i$'s ratings for all items, and $A_i' = C^{(i)}Q(Q^TC^{(i)}Q+\lambda_1U^TU+\lambda_2I)^{-1}$ and  $B_i'=\lambda_1U(Q^TC^{(i)}Q+\lambda_1U^TU+\lambda_2I)^{-1}$ are dependent on $r_i$.

Since $C^{(i)}$ is a function of $r_i$, the system consists of $l_x+m$ variables and $(l_x+m)\times K$ equations. Therefore, it's possible to recover the user information by solving the equation system. Similarly, a random factor $n \in [1,K]$ is fixed to align the number of equations and variables.

\subsection{Reconstruction Attack for \textit{SGD} Update}
For \textit{SGD}, the FL server is able to recover the user information within only two epochs given that the server has access to $U$ and $Q$.

\textbf{Attack for \textit{ObsOnly}:} After two epochs, the gradients FL server receives from users $i$ is given by:
\begin{equation} \label{gradSGDObsOnly}
\begin{gathered}
f^t(i,j)=(r_{i,j}-p^t_i{(q_j^t)}')p^t_i, j \in O_i\\
f^{t-1}(i,j)=(r_{i,j}-p^{t-1}_i{(q_j^{t-1})}')p^{t-1}_i, j \in O_i\\
f^t(i,d_u)=(x_{i,d_u}-p^t_i{(u^t_{d_u})}')p^t_i, d_u \in [1, l_x]\\
f^{t-1}(i,d_u)=(x_{i,d_u}-p^{t-1}_i{(u^{t-1}_{d_u})}')p^{t-1}_i, d_u \in [1, l_x]\\
\end{gathered}
\end{equation}
In pure SGD, the user latent factor is updated using equation (\ref{updatepSGD}) and (\ref{pgrad}). Plugging into the first gradient of equation (\ref{gradSGDObsOnly}), we have:
\begin{equation}\label{gradSGDObsOnly1}
\begin{gathered}
    f_n^t(i,j)=\bigl(r_{ij}-(p_i^{t-1}-\gamma\frac{\partial J}{\partial p_i^{t-1}})(q_j^{t-1}+\Delta q_j^t)'\bigr)\\\times (p_{in}^{t-1}-\gamma\frac{\partial J}{\partial p_{in}^{t-1}})
\end{gathered}
\end{equation}
where $\Delta q_j^t=q_j^t-q_j^{t-1}$, $f_n^t(i,j)$ denote the $n^{th}$ element of $f^t(i,j)$, and $p_{in}^{t-1}$ denote the $n^{th}$ element of $p_i^{t-1}$.

Equation (\ref{gradSGDObsOnly1}) is a multiplication of two terms. By looking at the first term, we have:
\begin{equation}\label{gradSGDObsOnly2}
\begin{gathered}
    r_{ij}-(p_i^{t-1}-\gamma\frac{\partial J}{\partial p_i^{t-1}})(q_j^{t-1}+\Delta q_j^t)'=r_{ij}\\-p_i^{t-1}{(q_j^{t-1})}'-p_i^{t-1}(\Delta q_j^t)' + \gamma\frac{\partial J}{\partial p_{in}^{t-1}} (q_j^t)'\\
    =\frac{G_n(j)}{p_{in}^{t-1}}+p_i^{t-1}g
\end{gathered}
\end{equation}
where:
\begin{equation}
\begin{gathered}
G_n(j)=f_n^{t-1}(i,j)-2\gamma(\sum_{k\in O_i}{f_n^{t-1}(i,k)q_k^{t-1}}\\
+\lambda_1\sum_{d_u}{f_n^{t-1}(i,d_u)u_{d_u}^{t-1}}){(q_j^t)}'\\
g=2\gamma\lambda_2{(q_j^t)}'-(\Delta q_j^t)'
\end{gathered}
\end{equation}
Then we look at the second term of equation (\ref{gradSGDObsOnly1}), which is given by:
\begin{equation}
\begin{gathered}
p_{in}^{t-1}-\gamma\frac{\partial J}{\partial p_{in}^{t-1}}=p_{in}^{t-1}(1-2\gamma\lambda_2)+\frac{2\gamma}{p_{in}^{t-1}}\\
\times [\sum_{k\in O_i}{f_n^{t-1}(i,k)q_{kn}^{t-1}}+\lambda_1\sum_{d_u}{f_n^{t-1}(i,d_u)u_{d_n}^{t-1}}]\\
=p_{in}^{t-1}(1-2\gamma\lambda_2)+\frac{F_n}{p_{in}^{t-1}}
\end{gathered}
\end{equation}
where:
\begin{equation}
\begin{gathered}
F_n=2\gamma[\sum_{k\in O_i}{f_n^{t-1}(i,k)q_{kn}^{t-1}}+\\
\lambda_1\sum_{d_u}{f_n^{t-1}(i,d_u)u_{d_un}^{t-1}}]
\end{gathered}
\end{equation}
Then equation (\ref{gradSGDObsOnly1}) can be written as:
\begin{equation}\label{SGDObsOnlyFianl}
\begin{gathered}
f_n^t\left(i,j\right)=\bigl(\frac{G_n(j)}{p_{in}^{t-1}}+p_i^{t-1}g\bigr)\bigl(p_{in}^{t-1}(1-2\gamma\lambda_2)+\frac{F_n}{p_{in}^{t-1}}\bigr)
\end{gathered}
\end{equation}
For $n \in [1,K]$, $j \in O_i$, where $p^{t-1}_i$ is the variable to solve. Noted that $G_n(j)$, $g$, and $F_n$ could be computed on the FL server.

Since there are $K$ variables and $K\times|O_i|$ equations, there should exist a solution $p^{t-1}_i$ satisfy the system (\ref{SGDObsOnlyFianl}). To reconcile the number of equations and variables, we choose a random item $j \in O_i$, and solve the equation systems under the fixed $j$.

After obtaining $p^{t-1}_i$, the server could compute $r_{i,j}$ and $x_{i,d_u}$ as followed:
\begin{equation}
\begin{gathered}
r_{ij}=\frac{f_n^{t-1}(i,j)}{p_{in}^{t-1}}+p_i^{t-1}{(q_j^{t-1})}'\\
x_{id_u}=\frac{f_n^{t-1}(i,d_u)}{p_{in}^{t-1}}+p_i^{t-1}{(u_{d_u}^{t-1})}'
\end{gathered}
\end{equation}

\textbf{Attack for \textit{InclUnc}:} Similarly, the FL server first obtain the equation system for $p^{t-1}_i$ given by:
\begin{equation}\label{SGDInclUncFinal}
\begin{gathered}
    f_n(i,j)=\bigl(\frac{f_n^{t-1}(i,j)}{p_{in}^{t-1}}+ \frac{c_{i,j}G_n'(j)}{p_{in}^{t-1}} + c_{i,j}p_i^{t-1}g' \bigr)\\
    \bigl(\frac{F_n'}{p_{in}^{t-1}}+p_{in}^{t-1}(1-2\gamma \lambda_2)\bigr), n \in [1,K], j \in [1,m]
\end{gathered}
\end{equation}
where:
\begin{equation}
\begin{gathered}
    G_n'(j)=-2\gamma(\sum_k{f_n^{t-1}(i,k)q_k^{t-1}}\\+\lambda_1\sum_{d_u}{f_n^{t-1}(i,d_u)u_{d_u}^{t-1}}){(q_j^t)}'\\
    F_n' = 2\gamma (\sum_k f_n^{t-1}(i,k)q_{kn}^{t-1} +\lambda_1\sum_{d_u}f_n^{t-1}(i,d_u)u_{d_u,n}^{t-1}\\
    g'=2\lambda_2 \gamma p_{j}^{t-1}-\Delta p_j^t
\end{gathered}
\end{equation}
For detail derivation of equation (\ref{SGDInclUncFinal}) refer to appendix \ref{attackderive}. Noted that $c_{i,j}$ is a function of $r_{i,j}$, which is dependent on $p^{t-1}_i$ based on equation (\ref{rating}). Therefore, $c_{i,j}$ is linked with $p^{t-1}_i$.

Given $K$ variables and $K\times m$ equations, the server should be able to find a solution $p^{t-1}_i$ for the system. Similarly, a random item $j \in [1, m]$ is fixed when solving the equation system.

Then the rating and user attributes could be computed as:
\begin{equation}\label{rating}
\begin{gathered}
r_{i,j} = \frac{f_n^{t-1}(i,j)p_i^t(q_j^t)'p_{in}^t-f_n^{t}(i,j)p_i^{t-1}(q_j^{t-1})'p_{in}^{t-1}}{f_n^{t-1}(i,j)p_{in}^t-f_n^{t}(i,j)p_{in}^{t-1}}\\
x_{i,d_u}=\frac{f_n^{t-1}(i,d_u)}{p_{in}^{t-1}}+p_i^{t-1}{(u_{d_u}^{t-1})}'
\end{gathered}
\end{equation}
where $p_{in}^t$ and $p_i^t$ can be obtained from formula (\ref{updatepSGD}) and (\ref{pgrad}).

\section{Privacy-Preserving MVMF (PriMVMF)}
\label{privmvmf}
To prevent information leakage, we develop \textit{PrivMVMF}, a privacy-preserving federated MVMF framework enhanced with homomorphic encryption (HE). In this framework, the client encrypts the gradients before sending them to the server, and the server can perform computation on the encoded gradients. The above attacks are based on access to individual gradients, while in HE, these gradients are sent to the server in encrypted form, rendering the reconstruction attacks infeasible.
\subsection{Paillier Cryptosystem}
This study utilized a partially HE scheme - Paillier cryptosystem \cite{paillier1999public}, which consists of three parts: key generation, encryption, and decryption.
\begin{itemize}
    \item \textit{Key generation}: Based on the $keysize$,  $(sk,pk)=\boldsymbol{Gen}(keysize)$ returns the public key $pk$ shared among all participants, and secret key $sk$ distributed only among the clients. Before the training process, one of the users generates a key pair.
    \item \textit{Encryption}: $c=\boldsymbol{Enc}(m,pk)$ encrypts message $m$ to cyphertext $c$ using public key $pk$.
    \item \textit{Decryption}: $m=\boldsymbol{Dec}(c,sk)$ reverses cyphertext $c$ to message $m$ using secret key $sk$.
\end{itemize}
Given two plaintexts $m_1$ and $m_2$, Paillier cryptosystem $E$ has the following properties:
\begin{itemize}
    \item \textit{Addition}: $E(m_1)\cdot E(m_2)=E(m_1+m_2)$.
    \item \textit{Multiplication}: ${E(m_1)}^{m_2}=E(m_1\cdot m_2)$
\end{itemize}

\textbf{Number Encoding Scheme:} Paillier encryption is only defined for non-negative integer, but the recommendation system contains float and negative numbers. The study follows Chai et al.’s method to convert floating points and negative numbers into unsigned integer \cite{chai2020secure}.

\textbf{Sampling of Unrated Item:} For the treatment of unrated item, this framework strikes a balance between efficiency and privacy protection. The \textit{ObsOnly} method is efficient while it reveals what items has been rated by the user. The \textit{InclUnc} method leaks no information but is computation intensive. 
To reconcile the two objectives, we design a strategy to randomly sample a portion of unrated items. Then the $c_{i,j}$ is given as followed:
\begin{equation}
c_{i,j}=\left\{
\begin{array}{ll}
1, & r_{i,j}>0\\
\alpha, & r_{i,j}=0\ and\ samp_{i,j}=1\\
0, & r_{i,j}=0\ and\ samp_{i,j}=0
\end{array}
\right.
\end{equation}
where $0<\alpha<1$, ${samp}_{i,j}=1$ if item $j$ appears in the sampled unrated items for user $i$, and ${samp}_{i,j}=0$ otherwise. Users only send the gradients with $c_{i,j}>0$.

For each user, we determine the number of sampled unrated items as a multiple of his rated items, denoted by $\rho$. Then the upper-bound probability that the FL server could correctly infer whether a given item is rated by the user is given by $\frac{1}{\rho+1}$.

\textbf{Decrypting Clients:} It's time-consuming to perform the update using the encrypted gradients. To reduce complexity, the server sends the aggregated gradient to some decrypting users for decryption, and uses the plaintext aggregated gradients to update the parameters. 

\textbf{Algorithms:} The detailed steps of \textit{PrivMVM} are shown in Algorithm \ref{alg:PriMVMF}. Noted that for the update of user latent factor $P$ and item feature latent factor $V$, we adopt the \textit{SemiALS} strategy for the following reason: although \textit{SemiALS} has higher time complexity per iteration, it requires fewer iterations to achieve the optimum and thus fewer encryption and decryption operations, the bottleneck of the HE scheme.

\textbf{Privacy Analysis}: The privacy of the algorithm is analyzed in terms of information leakage, which is characterized into two forms: i) original information, the observed user data , and ii) latent information, properties of user data \cite{mo2021quantifying}. We assume an honest-but-curious server for the analysis, i.e., the server will not deviate from the defined protocol but attempt to learn information from legitimately received messages. During the training of \textit{PrivMVMF}, the individual gradients are sent to the server in the encrypted form, and only the plaintext aggregated gradients are available to the server. The following shows that given the aggregated gradients, it leaks trivial original information about user data to the server.

Let $f(j)$, $f(d_u)$ be the aggregated gradients for item $j$ and user feature $d_u$, given by:
\begin{equation}
\begin{gathered}
    f(j)=\sum_i f(i,j) = \sum_{i:j\in O_i'} c_{i,j} (r_{ij}-p_iq_j^T)p_i, j\in [1,m]\\
    f(d_u)=\sum_i f(i,d_u) = \sum_i (x_{i,d_u}-p_iu_{d_u}^T)p_i, d_u\in [1,l_x]
\end{gathered}
\end{equation}
where $O_i'$ denotes the set of items rated by or appeared in the sampled unrated items for user $i$.

In \textit{PrivMVMF}, $p_i$ is updated by:
\begin{equation}
\label{gradprivmvmf}
    p_i = (r_i C^{(i)}Q_i' + \lambda_1 x_i U)((Q_i')^TC^{(i)}Q_i'+\lambda_1U^TU+\lambda_2 I)^{-1}
\end{equation}
where $Q_i'=Q_{|O_i'\times K|}$ is the latent factors for items in $O_i'$.

Let $A_i' = C^{(i)}Q_i'((Q_i')^TC^{(i)}Q_i'+\lambda_1U^TU+\lambda_2I)^{-1}$, and $B_i'=\lambda_1U((Q_i)'^TC^{(i)}Q_i'+\lambda_1U^TU+\lambda_2I)^{-1}$. Then $p_i$ can be written as:
\begin{equation}
\begin{gathered}
    p_i = r_iA_i'+x_iB_i'
\end{gathered}
\end{equation}

Plugging into equation (\ref{gradprivmvmf}), we can obtain the equation system as followed:
\begin{equation}
\begin{gathered}
    f(j) = \sum_i c_{i,j}(r_{ij}r_iA_i'+r_{ij}x_iB_i' -q_j((A_i')^T R_r A_i' \\
    +(A_i')^TR_xB_i'+(B_i')^TX_rA_i'+(B_i')^TX_xB_i')), j\in [1,m]\\
    f(d_u) = \sum_i (x_{i,d_u}r_iA_i'+x_{i,d_u}x_iB_i' -u_{d_u}((A_i')^T R_r A_i' \\+ (A_i')^TR_xB_i'+(B_i')^TX_rA_i'+(B_i')^TX_xB_i')), d_u\in [1,l_x]\\
\end{gathered}
\end{equation}
where:
\begin{equation}
\begin{gathered}
    R_r=r_i^Tr_i, R_x=r_i^Tx_i,   X_r=x^T_ir_i, X_x=x^T_ix_i
\end{gathered}
\end{equation}
The non-linear system consists of $(m+l_x)\times K$ equations and $\sum_iO_i'+nl_x$ variables. When $\sum_iO_i'+nl_x>>(m+l_x)\times K$, i.e., the user size is large enough, it's hard for the server to derive the original information of users.

\begin{algorithm}[htp]
   \caption{\textit{PrivMVM}}
   \label{alg:PriMVMF}
\begin{algorithmic}
\STATE Ramdomly select some clients as decrypters
\STATE \textbf{FL Server:}
\STATE \textbf{Initialize} $U$ and $Q$.
\FOR{t = 1 \textbf{to} T}{
\STATE Receive and aggregate encrypted $f(i,j)$ and $f(i,d_u)$ 
from user $i$ for $i\in[1,\ n]$. 
\STATE Send encrypted $\sum_i f(i,j)$ and $\sum_i f(i,d_u)$ 
to decrypters. 
\STATE Receive decrypted $\sum_i f(i,j)$ and $\sum_i f(i,d_u)$ 
from decrypters. 
\STATE Receive $f(j,d_v)$ from item server. 
\STATE Update $U$ using equation (\ref{updateU}).
\STATE Update $Q$ using equation (\ref{updateQ}).
}
\ENDFOR
\STATE
\STATE \textbf{Item Server:}
\WHILE{True}{
\STATE Receive $Q$ from FL server.
\STATE Compute local $V$ using equation (\ref{updateV}).
\STATE Compute item latent factor gradients
$f(j,d_v)$.
\STATE Transmit gradients to server.
}
\ENDWHILE
\STATE
\STATE \textbf{Client:}
\WHILE{True}{
\STATE Receive $U$ and $Q$ from server.
\STATE Compute local $p_i$ using equation (\ref{pals}).
\STATE Compute $U$ gradients
$f(i,d_u)$ for $d_u\in[1,l_x]$.
\STATE Compute $Q$ gradients
$f(i,j)$ for $j\in[1,m]$.
\STATE Transmit gradients to server.
}
\ENDWHILE
\STATE
\STATE \textbf{Decrypter:}
\WHILE{True}{
\STATE Receive encoded $\sum_i f(i,j)$ and $\sum_i f(i,d_u)$ 
from FL server. 
\STATE Decrypt and transmit $\sum_i f(i,j)$ and $\sum_i f(i,d_u)$ 
to FL server. 
}
\ENDWHILE
\end{algorithmic}
\end{algorithm}

\section{Experiments}
\label{experiment}
\subsection{Dataset and Experimental Setup}
The experiment is performed on MovieLens-1M dataset\footnote{https://grouplens.org/datasets/movielens/1m/}. The dataset contains 914676 ratings from 6040 users on 3952 movies, with each user submitting at least 20 ratings. The experiment is implemented on Ubuntu Linux 20.04 server with 32-core CPU and 128GB RAM, where the programming language is Python. 

We construct the rating matrix based on the explicit ratings, where the missing values are set to zero. The following user attributes are considered: \textit{Age}, \textit{Gender}, \textit{Occupation} and \textit{Zipcode}. \textit{Age} is discretized into seven groups with equal interval, and \textit{Zipcode} is linked to the US region. The movie features are described by the tag genome dataset containing $1,128$ tags for $9,734$ movies. To reduce dimensionality, we take the first 20 principal components for the tags features.

We use Bayesian optimization \cite{snoek2012practical} approach based on four-fold cross validation to optimize the hyperparameters. Table \ref{hyper} summarizes the hyperparameters for the experiment.

\begin{table}[ht]
\caption{Hyperparameter for MVMF on Movielens Dataset}
\label{hyper}
\tablefont%
\setlength{\tabcolsep}{3pt}
\begin{tabular*}{21pc}{@{\extracolsep{\textwidth minus \textwidth}}lcccccc@{}}
\hline
 \textbf{Hyperparameter} & $K$ & $\beta_{1}$ & $\beta_{2}$ & $\varepsilon$ & $\gamma$ & $\alpha$\\
\hline
 \textbf{Value} & $6$ & $0.5$ & $0.99$ & $1e-8$ & $0.05$ & $0.1$\\
\hline
 \textbf{Hyperparameter} & $\lambda_{1}$ & $\lambda_{2}$ & $\rho$ & $|iter|$ & $|epo|$ & $l_p$\\
\hline
 \textbf{Value} & $1$ & $10$ & $1$ & $10$ & $20$ & $1024$ \\
\hline
\multicolumn{7}{l}{}\\[-5pt]
\multicolumn{7}{@{\extracolsep{\fill}}p{21pc}@{\extracolsep{\fill}}}{\hspace*{9pt} $|iter|$ and $|epo|$ denote the number of iterations and epochs. $\rho$ is the proportion of sampled items based on rated items. $l_p$ denote the length of public key.}
\end{tabular*}
\end{table}

\subsection{Server Attack}
\textbf{Solving Nonlinear System:} To perform a server reconstruction attack, we first developed the equation systems described in section \ref{serveattack}. To solve the non-linear systems, we experiment with the four methods\cite{osti_6997568, broyden1965class, la2006spectral, anderson1965iterative}: modified Powell's hybrid method, Broyden’s bad method, Scalar Jacobian approximation, and Anderson mixing, and select the best method within each scenario using a sample of 100 users. Refer to table \ref{accdiffmethod} for the selected method within each scenario. 

\textbf{Smoothing $c_{i,j}$ as a Function of $r_{i,j}$ for \textit{InclUnc}:} In \textit{InclUnc}, $c_{i,j}=f(r_{i,j})$ is not a continuous function, while Jacobian matrix is needed for most of the iterative methods. To smooth $f$, we design the following function:
\begin{equation}
c_{i,j}=\left\{
\begin{array}{ll}
1, & r_{i,j}>1\\
r_{i,j}, & 1 \ge r_{i,j} > 0\\
0, & 0 \ge r_{i,j}
\end{array}
\right.
\end{equation}

\textbf{Evaluation metrics:} We employ accuracy to measure the performance of server inference, with steps as follows. After obtaining the estimation of $\hat{r}(i)$ and $\hat{x}_i$ for each user $i$, we clipped $\hat{r}(i)$ within $[0, R_{max}]$ and $\hat{x}_i$ within $[0, 1]$ ($x_i$ are all dummy variables), and then rounded the estimations to the nearest integers. The accuracy for user ratings and attributes is computed as follows:
\begin{equation}
\begin{gathered}
    Accuracy\ for\ Rating=\frac{1}{|usr|} \sum_i \frac{|\hat{r}_i=r_i|}{|r_i|}\\
    Accuracy\ for\ Attribute=\frac{1}{|usr|} \sum_i \frac{|\hat{x}_i=x_i|}{|x_i|}
\end{gathered}
\end{equation}
where $usr$ denotes the set of all users, and $\hat{r}_i$ and $\hat{x}_i$ denote the transformed estimation of $r_i$ and $x_i$.

\textbf{Result and Analysis:} Table \ref{accSRA} reports the accuracy of server reconstruction attack in four scenarios, from which we can make the following observations: (1) In all cases, the server is able to recover the user's private information with accuracy $> 80\%$, which is a non-negligible privacy concern. (2) For both \textit{SemiALS} and \textit{SGD}, including an uncertainty coefficient deteriorates the performance of server attack. One explanation is that $c_{i,j}$ is not a differentiable function of $r_{i,j}$, posing a challenge to obtaining the Jacobian matrix of the system. (3) Using \textit{SGD} method to update $P$ makes it harder for the server to infer user information given by the reduced accuracy.

\begin{table}[ht]
\caption{Accuracy of Server Reconstruction Attack}
\label{accSRA}
\begin{tabular*}{21pc}{@{\extracolsep{\textwidth minus \textwidth}}lcccr@{}}
\hline
 & \multicolumn{2}{c}{\textit{SemiALS}} & \multicolumn{2}{c}{\textit{SGD}} \\
  & \textit{ObsOnly} & \textit{InclUnc} & \textit{ObsOnly} & \textit{InclUnc} \\
\hline
Rating    & $0.9911$ & $0.8785$ & $0.8223$ & $0.8182$\\
Attribute & $0.9991$ & $0.8898$ & $0.9230$ & $0.8474$\\
\hline
\end{tabular*}
\end{table}

\textbf{Robustness Check:} We consider the case when a small amount of noise is added to the gradients. With perturbed gradients, the equation systems are solved using the following steps:
\begin{itemize}
    \item For \textit{SemiALS} (\textit{SGD} with \textit{ObsOnly} / \textit{SGD} with \textit{InclUnc}), compute the set of solutions for each $n\in [1,K]$ ($j \in O_i$ / $j$ from a randomly chosen set of items). 
    \item Clip the solutions to the correct domain.
    \item Take the average of the solutions and round the estimations to integers.
\end{itemize}

\begin{figure}
\centerline{\includegraphics[width=20pc]{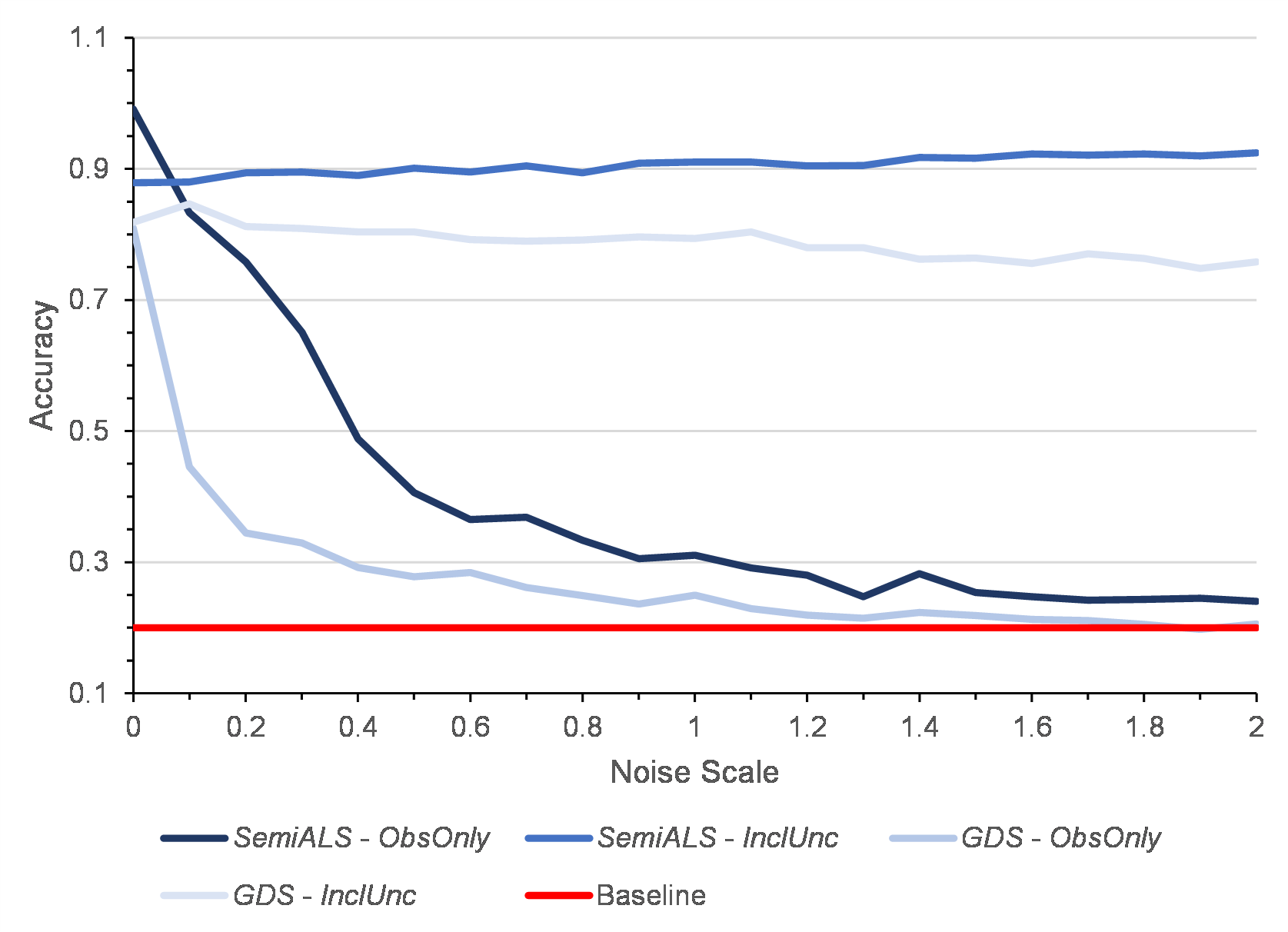}}
\caption{Accuracy for ratings under four cases after adding Laplace noises from $0$ to $2$.}
\label{rating_acc_four}
\end{figure}

\begin{figure}
\centerline{\includegraphics[width=20pc]{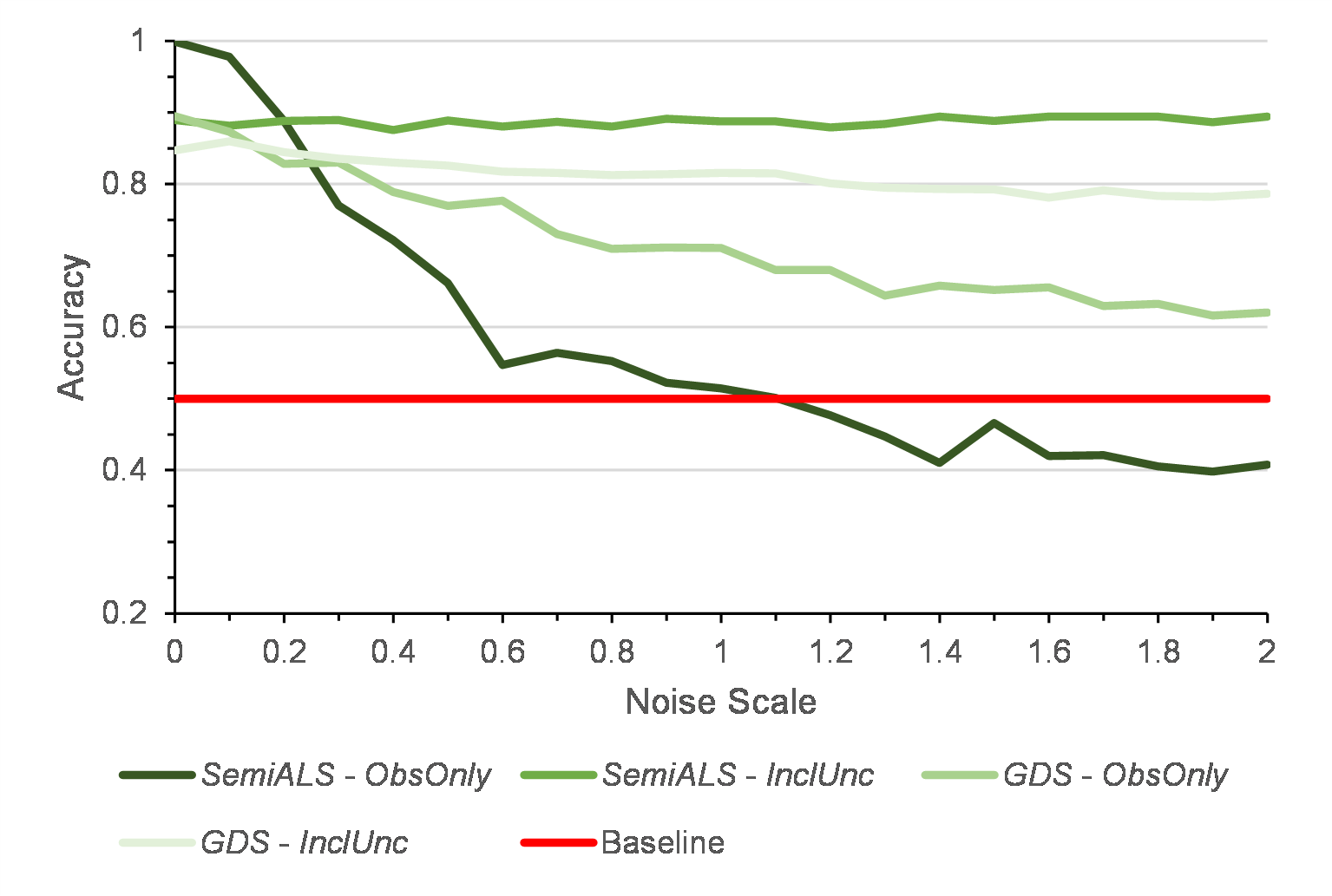}}
\caption{Accuracy for user attributes under four cases after adding Laplace noises from $0$ to $2$.}
\label{attribute_acc_four}
\end{figure}

We conduct the experiment under the four scenarios, where Laplace noises are added to the gradients with scale $b$ ranging from $0$ to $2$. The performance is given by fig. \ref{rating_acc_four} and fig. \ref{attribute_acc_four}. Baseline refers to the theoretical accuracy under the random guessing strategy. 

For both \textit{SemiALS} and \textit{SGD}, the reconstruction accuracy are stable and above the baseline by more than 50\% under the \textit{InclUnc} scenario. Refer to fig. \ref{acc_two} in appendix for the accuracy with noise scale up to $10^4$. For \textit{ObsOnly}, the rating accuracy approaches the baseline under both \textit{SemiALS} and \textit{SGD} when the noise scale increases to $2$, and the user attribute accuracy is below the baseline under \textit{SemiALS} when the noise scale exceeds $1$. In all scenarios, the attack accuracy is more than 30\% above the random guessing baseline for noise scale less than $0.5$.

The results reveal the following: (1) The reconstruction attack is effective under small amount of noises ($b\leq 0.5$). (2) Although including an uncertainty coefficient hides the unrated ratings, it makes the attack more resilient to random noises.

\subsection{\textit{PrivMVMF}}
\textbf{Evaluation metrics:} The study adopted the following four evaluation metrics: Normalized Discounted Cumulative Gain at 10 (NDCG@10), Precision, Recall, and F1. The accurate prediction is defined as an item recommended rated above a threshold by the given user\cite{bobadilla2013recommender}. Refer to appendix \ref{metricsprivmvmf} for the details of evaluation metrics.

\textbf{Scenarios of Testing:} Both approaches can provide recommendations for new users and items. Table \ref{accHE} presents the performance of the three scenarios: \textit{Existing User and Item}, \textit{Cold-start Item}, and \textit{Cold-start User}. For \textit{Existing User and Item}, the items of each user are randomly divided into 80\% training set and 20\% testing set. The items and users in the testing set are supposed to have a rating history. For \textit{Cold-start Item}, the items are randomly divided into 90\% training set and 10\% testing set. The testing items are treated as new items without a rating history. For \textit{Cold-start User}, a random subset of 10\% users is held out as new users for testing.
\begin{table}[ht]
\caption{Test Accuracy of \textit{FedMVMF} and \textit{PrivMVMF}}
\label{accHE}
\begin{tabular*}{21pc}{@{\extracolsep{\textwidth minus \textwidth}}p {4pc}|p{4.8pc}|p{4.8pc}|p{4.5pc} @{}}
\hline
 & \multicolumn{1}{c|}{\textit{FedMVMF}} & \multicolumn{1}{c|}{\textit{PrivMVMF}} & \multicolumn{1}{c}{Diff\%} \\
\hline
& \multicolumn{3}{c}{Existing User and Item} \\
\hline
NDCG@10 & \multicolumn{1}{c|}{$0.8304$ \tiny$\pm 0.0008$\normalsize} & \multicolumn{1}{c|}{$0.8313$ \tiny$\pm 0.0003$\normalsize} & \multicolumn{1}{c}{$0.11$}
 \\
\hline
Precision & \multicolumn{1}{c|}{$0.2677$ \tiny$\pm 0.0034$\normalsize} & \multicolumn{1}{c|}{$0.2931$ \tiny$\pm 0.0047$\normalsize} & \multicolumn{1}{c}{$8.69$} \\
\hline
Recall & \multicolumn{1}{c|}{$0.2092$ \tiny$\pm 0.0048$\normalsize} & \multicolumn{1}{c|}{$0.2208$ \tiny$\pm 0.0034$\normalsize} & \multicolumn{1}{c}{$5.24$} \\
\hline
F1 & \multicolumn{1}{c|}{$0.2348$ \tiny$\pm 0.0043$\normalsize} & \multicolumn{1}{c|}{$0.2519$ \tiny$\pm 0.0040$\normalsize} & \multicolumn{1}{c}{$6.76$}\\
\hline
& \multicolumn{3}{c}{Cold-start User} \\
\hline
NDCG@10 & \multicolumn{1}{c|}{$0.7060$ \tiny$\pm 0.0042$\normalsize} & \multicolumn{1}{c|}{$0.7090$ \tiny$\pm 0.0092$\normalsize} & \multicolumn{1}{c}{$0.42$}\\
\hline
Precision & \multicolumn{1}{c|}{$0.2668$ \tiny$\pm 0.0076$\normalsize} & \multicolumn{1}{c|}{$0.2728$ \tiny$\pm 0.0081$\normalsize} & \multicolumn{1}{c}{$2.18$}\\
\hline
Recall & \multicolumn{1}{c|}{$0.0404$ \tiny$\pm 0.0010$\normalsize} & \multicolumn{1}{c|}{$0.0423$ \tiny$\pm 0.0015$\normalsize} & \multicolumn{1}{c}{$4.66$}\\
\hline
F1 & \multicolumn{1}{c|}{$0.0701$ \tiny$\pm 0.0017$\normalsize} & \multicolumn{1}{c|}{$0.0733$ \tiny$\pm 0.0024$\normalsize} & \multicolumn{1}{c}{$4.33$}\\
\hline
& \multicolumn{3}{c}{Cold-start Item} \\
\hline
NDCG@10 & \multicolumn{1}{c|}{$0.8208$ \tiny$\pm 0.0028$\normalsize} & $0.8193$ \tiny$\pm 0.0069$\normalsize & \multicolumn{1}{c}{$0.18$}\\
\hline
Precision & \multicolumn{1}{c|}{$0.1781$ \tiny$\pm 0.0077$\normalsize} & $0.1735$ \tiny$\pm 0.0082$\normalsize & \multicolumn{1}{c}{$2.60$}\\
\hline
Recall & \multicolumn{1}{c|}{$0.2497$ \tiny$\pm 0.0114$\normalsize} & $0.2563$ \tiny$\pm 0.0289$\normalsize & \multicolumn{1}{c}{$2.56$}\\
\hline
F1 & \multicolumn{1}{c|}{$0.2075$ \tiny$\pm 0.0021$\normalsize} & $0.2064$ \tiny$\pm 0.0127$\normalsize & \multicolumn{1}{c}{$0.55$}\\
\hline
\multicolumn{4}{l}{}\\[-5pt]
\multicolumn{4}{@{\extracolsep{\fill}}p{21pc}@{\extracolsep{\fill}}}{\hspace*{9pt} The values denote the $mean \pm standard\ deviation$ of the performance.}
\end{tabular*}
\end{table}

\textbf{Accuracy:} Table \ref{accHE} compares the testing accuracy between \textit{FedMVMF} and \textit{PrivMVMF}, with each approach running for 5 rounds. Noted that the \textit{FedMVMF} adopts the same sampling strategy as \textit{PrivMVMF} for consistency, slightly different from that in section \ref{fedmvmf}. It can be observed that the difference in testing accuracy is trivial in all scenarios, suggesting that the proposed framework is lossless.

\textbf{Efficiency:} The model training in each stage can be divided into four phases: local update, aggregation, decryption, and server update. The study evaluates the time consumption in these four phases respectively.
\begin{itemize}
    \item \textit{Local update:} Clients compute the gradients and encryption them with the public key.
    \item \textit{Aggregation:} Server receives and aggregates the encrypted gradients from clients.
    \item \textit{Decryption:} Decrypting clients decrypts the aggregated gradients.
    \item \textit{Server update:} Server updates the latent factor matrix using decrypted aggregated gradients.
\end{itemize}
Table \ref{timeHE} presents the computation time in each epoch. It can be observed that the aggregation and decryption process take up most of the time. It can improve efficiency if the decryption workload is distributed to several clients instead of only one user. More work can be done to reduce the complexity of the operation of the encrypted gradient as well.
\begin{table}[ht]
\caption{Time Consumption in Each Phase (seconds)}
\label{timeHE}
\begin{tabular*}{21pc}{@{\extracolsep{\textwidth minus \textwidth}}lcc}
\hline
  & \textit{FedMVMF} & \textit{PrivMVMF} \\
\hline
\textbf{Local Update} & $0.0195$\tiny$\pm 0.0003$\normalsize & $3.0787$\tiny$\pm 0.0302$\\
\textbf{Aggregation} & $3.2878$\tiny$\pm 0.2563$\normalsize & $239.2998$\tiny$\pm 4.7052$\\
\textbf{Decryption} & / & $97.5748$\tiny$\pm 0.9483$\\
\textbf{Server Update} & $0.1912$\tiny$\pm 0.0672$\normalsize & $0.3334$\tiny$\pm 0.0333$\normalsize\\
\hline
\multicolumn{3}{l}{}\\[-5pt]
\multicolumn{3}{@{\extracolsep{\fill}}p{21pc}@{\extracolsep{\fill}}}{\hspace*{9pt} It assumes that there are one decrypter for \textit{PrivMVMF}. \textit{Local Update} represents the time spent in the phase per user. The \textit{FedMVMF} doesn’t have Decryption phase, so the value is “/”.}
\end{tabular*}
\end{table}

\section{Conclusions and future work}
\label{conclusion}
To understand the privacy risks in federated MVMF recommender systems, this paper provides a theoretical analysis of the server reconstruction attack in four scenarios. It also proposes \textit{PrivMVMF}, a privacy-preserving federated MVMF framework enhanced with HE, to overcome the information leakage problem. Empirical studies on MovieLens-1M dataset show that: (1) In \textit{FedMVMF}, the FL server could infer users' rating and attribute with accuracy $>80\%$ using plaintext gradients. (2) For \textit{OnbOnly}, the reconstruction attack is effective under Laplace noise with $b\leq 0.5$; for \textit{InclUnc}, it is effective with noise $b\leq 10$. (3) \textit{PrivMVMF} can protect user privacy well compared wth \textit{FedMVMF}. (4) Aggregation and decryption process occupy most of the time in \textit{PrivMVMF}.

Future work involves the following directions. Firstly, communication time could be investigated in \textit{PrivMVMF} framework. Secondly, it's interesting to improve the efficiency of HE since it's time-consuming to perform the operation on the encrypted gradients.

\bibliographystyle{IEEEtran}
\bibliography{ref}

\begin{appendix}
\subsection{Attack for \textit{SGD}-\textit{InclUnc}}
\label{attackderive}
After two epochs, the gradients FL server receives from user $i$ is given by:
\begin{equation} \label{gradSGDInclUnc}
\begin{gathered}
f^t(i,j)=c_{i,j}(r_{i,j}-p^t_i{(q_j^t)}')p^t_i, j \in [1, m]\\
f^{t-1}(i,j)=c_{i,j}(r_{i,j}-p^{t-1}_i{(q_j^{t-1})}')p^{t-1}_i, j \in [1, m]\\
f^t(i,d_u)=(x_{i,d_u}-p^t_i{(u^t_{d_u})}')p^t_i, d_u \in [1, l_x]\\
f^{t-1}(i,d_u)=(x_{i,d_u}-p^{t-1}_i{(u^{t-1}_{d_u})}')p^{t-1}_i, d_u \in [1, l_x]\\
\end{gathered}
\end{equation}
In \textit{SGD}, the user latent factor is updated using equation (\ref{updatepSGD}) and (\ref{pgrad}). Plugging into the first gradient of equation (\ref{gradSGDInclUnc}), we have:
\begin{equation}\label{gradSGDInclUnc1}
\begin{gathered}
    f_n^t(i,j)=c_{i,j}\bigl(r_{ij}-(p_i^{t-1}-\gamma\frac{\partial J}{\partial p_i^{t-1}})(q_j^{t-1}+\Delta q_j^t)'\bigr)\\
    \times (p_{in}^{t-1}-\gamma\frac{\partial J}{\partial p_{in}^{t-1}})
\end{gathered}
\end{equation}

Equation (\ref{gradSGDInclUnc1}) is a multiplication of two terms. By looking at the first term, we have:
\begin{equation}\label{gradSGDInclUnc12}
\begin{gathered}
    c_{i,j}\bigl(r_{ij}-(p_i^{t-1}-\gamma\frac{\partial J}{\partial p_i^{t-1}})(q_j^{t-1}+\Delta q_j^t)'\bigr)\\
    =\bigl(\frac{f_n^{t-1}(i,j)}{p_{in}^{t-1}}+ \frac{c_{i,j}G_n'(j)}{p_{in}^{t-1}} + c_{i,j}p_i^{t-1}g' \bigr)
\end{gathered}
\end{equation}
where:
\begin{equation}
\begin{gathered}
    G_n'(j)=-2\gamma(\sum_k{f_n^{t-1}(i,k)q_k^{t-1}}\\
    +\lambda_1\sum_{d_u}{f_n^{t-1}(i,d_u)u_{d_u}^{t-1}}){(q_j^t)}'\\
    g'=2\lambda_2 \gamma p_{j}^{t-1}-\Delta p_j^t
\end{gathered}
\end{equation}

The second term of equation (\ref{gradSGDInclUnc1}) is given by:
\begin{equation}
\begin{gathered}
p_{in}^{t-1}-\gamma\frac{\partial J}{\partial p_{in}^{t-1}}=\frac{F_n'}{p_{in}^{t-1}}+p_{in}^{t-1}(1-2\gamma \lambda_2)
\end{gathered}
\end{equation}
where:
\begin{equation}
\begin{gathered}
F_n' = 2\gamma (\sum_k f_n^{t-1}(i,k)q_{kn}^{t-1} +\lambda_1\sum_{d_u}f_n^{t-1}(i,d_u)u_{d_u,n}^{t-1}\\
\end{gathered}
\end{equation}
Then the multiplication gives equation system (\ref{SGDInclUncFinal}).

\subsection{Evaluation Metrics for \textit{PrivMVMF}}
\label{metricsprivmvmf}
The study sets the rating threshold to be 4, and the number of items recommended to be 10 per user. The metrics are defined as:
\begin{equation}
    NDCG@10=\frac{DCG@10}{iDCG@10}
\end{equation}
\begin{equation}
    Precision=\frac{1}{|usr|}\sum_{i}\frac{t_p^i}{t_p^u+f_p^i}
\end{equation}
\begin{equation}
    Recall=\frac{1}{|usr|}\sum_{i}\frac{t_p^i}{t_p^i+f_n^i}
\end{equation}
\begin{equation}
    F1=\frac{2\times Precision\times Recall}{Precision + Recall}
\end{equation}
where $t_p^i$ denotes the true positive for user $i$, $f_p^i$ denotes the false positive for user $i$, $f_n^i$ denotes the false negative for user $i$, $iDCG@10$ is the maximum possible $DCG@10$, and $DCG@10$ is given by:
\begin{equation}
    DCG@10=\sum_{i=1}^5\frac{2^{x_i-1}}{\log_2(i+1)}
\end{equation}
where $i$ is the item with the $i^{th}$ highest predicted rating, and $x_i$ is the actual rating for the item received by a given user.


\subsection{Table and Figures}
\begin{table}[ht]
\caption{Accuracy for attack with four methods}
\label{accdiffmethod}
\tablefont%
\setlength{\tabcolsep}{3pt}
\begin{tabular*}{21pc}{@{\extracolsep{\textwidth minus \textwidth}}lcccccc@{}}
\hline
 &  & \multicolumn{2}{c}{\textit{SemiALS}} & \multicolumn{2}{c}{\textit{SGD}}\\
 &  & \textit{ObsOnly} & \textit{InclUnc} & \textit{ObsOnly} & \textit{InclUnc}\\
\hline
\multirow{2}{*}{Hybr}& Rating & $\boldsymbol{0.91}$ & / & $0.60$ & $\boldsymbol{0.82}$ \\
& Attribute & $\boldsymbol{0.99}$ & / & $0.86$ & $\boldsymbol{0.84}$ \\
\hline
\multirow{2}{*}{Broyden}& Rating & $0.91$ & $0.53$ & $0.73$ & $0.71$ \\
& Attribute & $0.94$ & $0.60$ & $0.85$ & $0.73$ \\
\hline
\multirow{2}{*}{Scalar}& Rating & $0.01$ & $\boldsymbol{0.89}$ & $0.66$ & $0.24$\\
& Attribute & $0.35$ & $\boldsymbol{0.89}$ & $0.74$ & $0.42$ \\
\hline
\multirow{2}{*}{Anderson}& Rating & $0.55$ & $0.49$ & $\boldsymbol{0.77}$ & $0.57$ \\
& Attribute & $0.78$ & $0.46$ & $\boldsymbol{0.88}$ & $0.62$ \\
\hline
\multicolumn{6}{l}{}\\[-5pt]
\multicolumn{6}{@{\extracolsep{\fill}}p{21pc}@{\extracolsep{\fill}}}{\hspace*{9pt} Hybr, Broyden, Scalar, and Anderson denote modified Powell’s hybrid method, Broyden's bad method, Scalar Jacobian approximation and Anderson mixing respectively. For \textit{SemiALS} update with \textit{InclUnc}, Hybr doesn't return the solution so the value is "/". The selected method for each scenario is marked with bold face.}
\end{tabular*}
\end{table}

\begin{figure}[hbp]
\subfloat[ratting]{%
\includegraphics[clip,width=18pc]{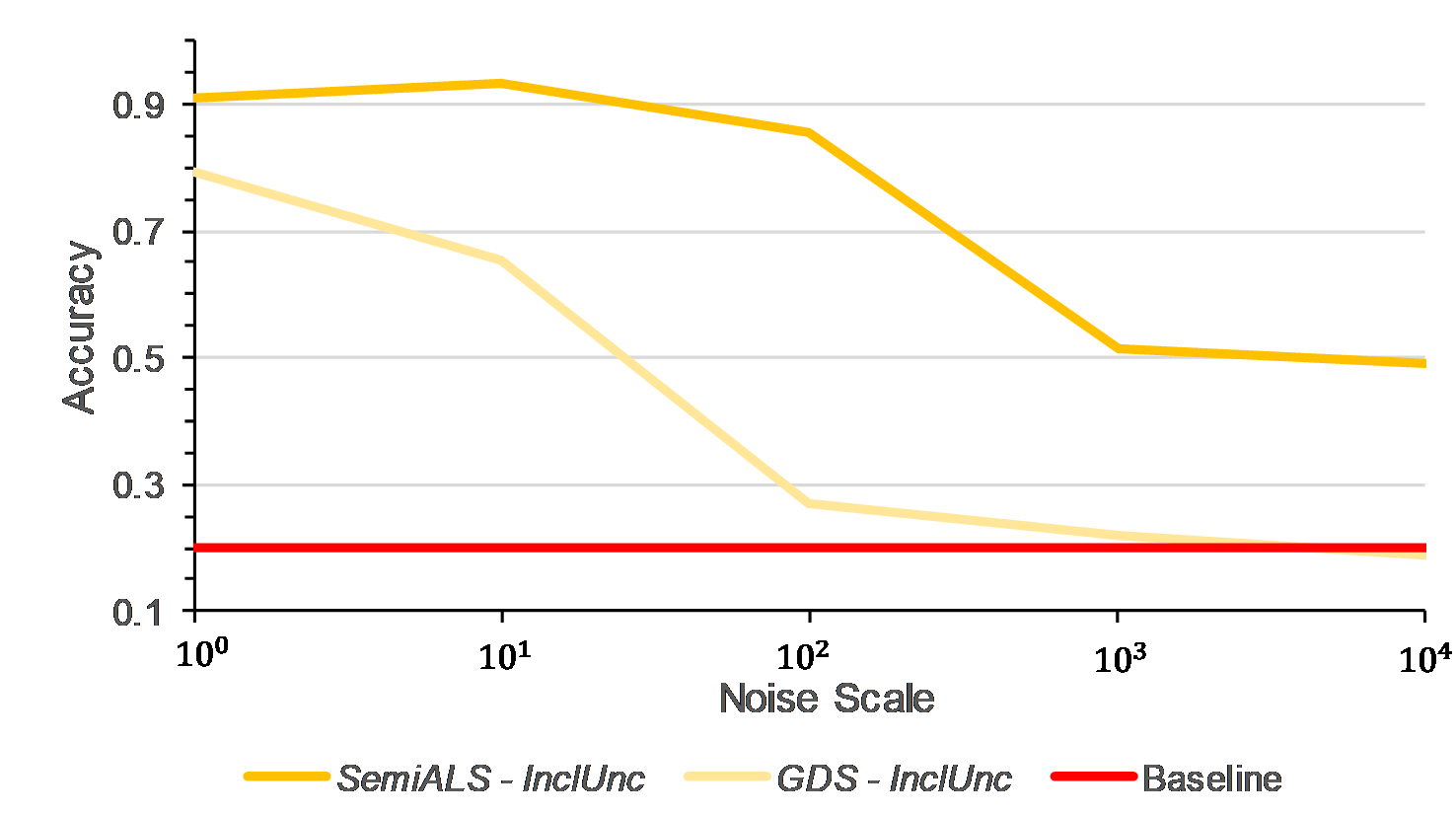}
}\\
\subfloat[user attributes]{%
\includegraphics[clip,width=18pc]{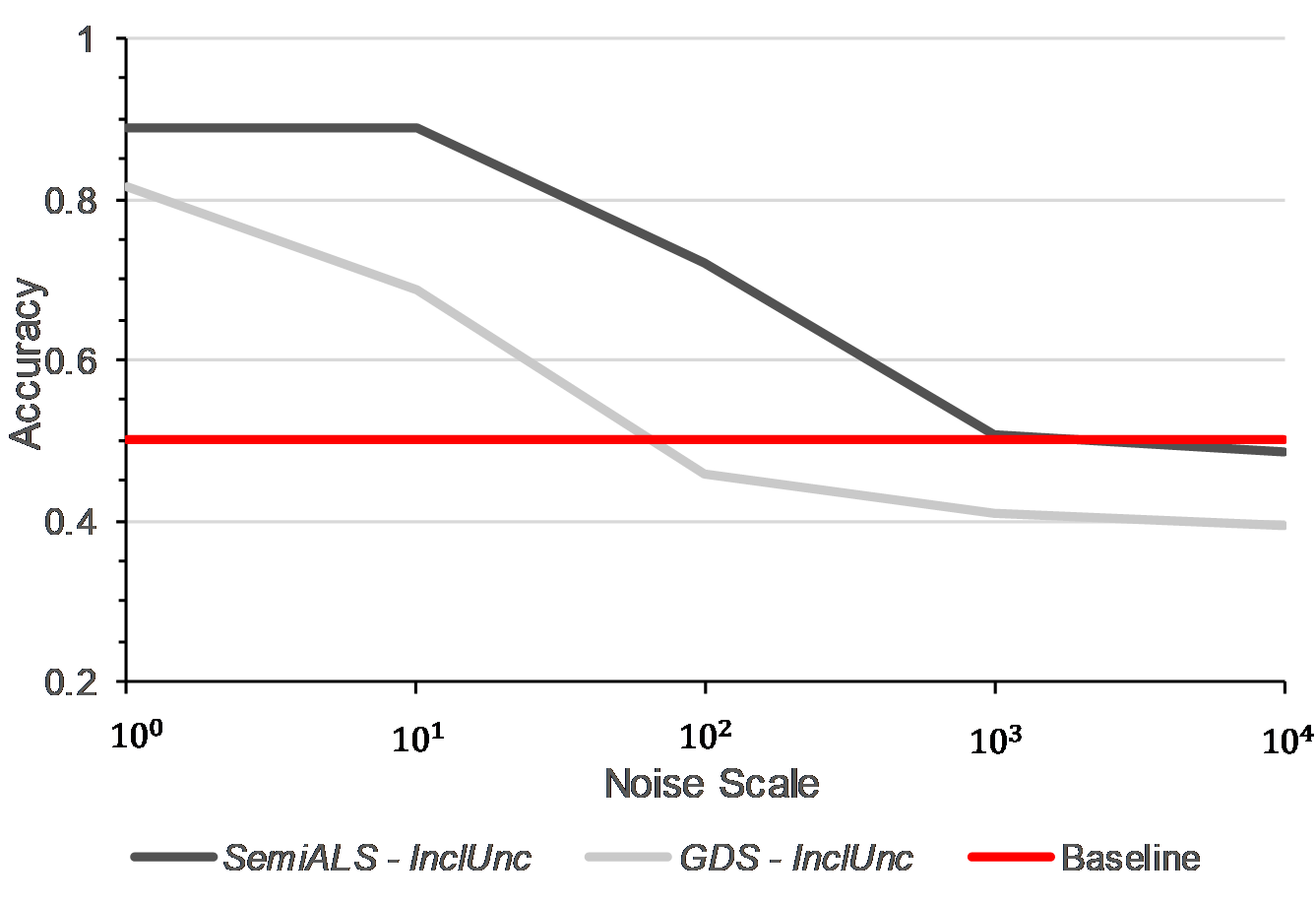}
}
\caption{Accuracy two \textit{InclUnc} cases after adding Laplace noises from $1$ to $10000$.}
\label{acc_two}
\end{figure}
\end{appendix}

\end{document}